\documentclass[preprint,aps,tightenlines,showpacs,superscriptaddress]{revtex4}
\usepackage[dvips]{graphicx}
\usepackage{dcolumn}
\usepackage{epsfig}

\begin{document}
\title{Extraction of Transition Form Factors for Nucleon Resonances within a
Coupled-Channels Model\footnote{Notice: Authored by
Jefferson Science Associates, LLC under U.S. DOE Contract No. DE-AC05-06OR23177. 
The U.S. Government retains a non-exclusive, paid-up, irrevocable, world-wide
license to publish or reproduce this manuscript for U.S. Government purposes.
} }
\author{N. Suzuki}
\affiliation{Department of Physics, Osaka University, Toyonaka,
Osaka 560-0043, Japan}
\affiliation{ Excited Baryon Analysis Center (EBAC), Thomas Jefferson National
Accelerator Facility, Newport News, VA 23606, USA}
\author{T. Sato}
\affiliation{Department of Physics, Osaka University, Toyonaka,
Osaka 560-0043, Japan}
\affiliation{ Excited Baryon Analysis Center (EBAC), Thomas Jefferson National
Accelerator Facility, Newport News, VA 23606, USA}
\author{T.-S. H. Lee}
\affiliation{Physics Division, Argonne National Laboratory,
Argonne, IL 60439, USA}
\affiliation{ Excited Baryon Analysis Center (EBAC), Thomas Jefferson National
Accelerator Facility, Newport News, VA 23606, USA}

\begin{abstract}
We explain how an analytic continuation we have developed recently 
is applied to determine
the residues of the nucleon resonance poles within a dynamical
coupled-channel model of meson-baron reactions. 
A procedure for evaluating the electromagnetic $N$-$N^*$ transition
form factors at resonance poles is developed.
Illustrative 
results of the obtained $N^*\rightarrow \pi N, \gamma N$ transition form factors
for $P_{11}$, $P_{33}$, and $D_{13}$ nucleon resonances are presented and
compared with previous results.

\end{abstract}

\maketitle

\newpage

\section{Introduction}

In a recent paper\cite{ssl09}, we have developed an analytic continuation
method to determine nucleon resonances within a dynamical coupled-channel
model of meson-baryon reactions\cite{msl07} (MSL). The method has been 
applied\cite{sjklms09} to extract 14 nucleon resonances from the $\pi N$ model
developed in Ref.\cite{jlms07} (JLMS) which has been extended to
investigate $\pi N, \gamma N\rightarrow \pi \pi N$\cite{kjlms09}, 
$\gamma N \rightarrow \pi N$\cite{jlmss08} and $N(e,e'\pi) N$\cite{jklmss09}
 reactions.
The purpose of this paper is to explain how the residues of the 
extracted resonance poles are determined from the    
predicted $\pi N \rightarrow \pi N$ and
$\gamma^* N \rightarrow \pi N$  amplitudes.

In section II, we briefly review the analytic continuation method developed
in Ref.\cite{ssl09}. Section III is devoted to explaining the determination
of the residues of the nucleon resonance poles. Illustrative results for
$P_{11}$, $P_{33}$, and $D_{13}$ nucleon resonances are presented
in section IV
and compared with the results from other analysis.
A summary is given in section V.
\section{Analytic continuation method}

Within the  MSL formulation\cite{msl07},
the partial wave amplitudes of 
two-body meson-baryon reactions
can be written 
as
\begin{eqnarray}
T_{\beta,\alpha}(p',p;E)
& = & t_{\beta,\alpha}(p',p;E)
   + t^R_{\beta,\alpha}(p',p;E) \,,
\label{eq:fullt}
\end{eqnarray}
where $\alpha, \beta$ represent the  meson-baryon (MB)
 states $\gamma N$, $\pi N,\eta N,\rho N, \sigma N, \pi\Delta$, and
\begin{eqnarray}
t^R_{\beta,\alpha}(p',p;E) &=& 
\sum_{i,j}\bar{\Gamma}_{\beta,i}(p';E)
[G_{N^*}(E)]_{i,j}
\bar{\Gamma}_{\alpha,j}(p;E) 
\label{eq:tr}
\end{eqnarray}
with
\begin{eqnarray}
[G_{N^*}^{-1}]_{i,j}(E)&=& 
(E - m_{N^*_i})\delta_{i,j} - \Sigma_{i,j} (E)\,.
\label{eq:tr-g}
\end{eqnarray}
Here $i,j$ denote the bare $N^*$ states defined in the Hamiltonian.
$ m_{N^*_i}$ are their masses.
The first term (called meson-exchange amplitude from nowon)
in Eq.(\ref{eq:fullt}) 
are defined by the following
equation
\begin{eqnarray}
t_{\beta,\alpha}(p',p;E)
& = & v_{\beta,\alpha}(p',p)
   +
\int_C dq q^2 \sum_{\gamma}
v_{\beta,\gamma}(p',q;E)G_\gamma(q,E)
t_{\gamma,\alpha}(q,p;E)
\label{eq:mext}
\end{eqnarray}
where $v_{\beta,\alpha}$ is defined by the meson-exchange mechanisms, and
$G_\gamma(q,E)$ is the propagator for channel $\gamma$.
The dressed vertices and the energy shifts of the second term in 
Eqs.(\ref{eq:tr})-(\ref{eq:tr-g}) are defined by
\begin{eqnarray}
\bar{\Gamma}_{\alpha,j}(p;E) & = &
\Gamma_{\alpha,j}(p) + \int_C dq q^2 \sum_\gamma
t_{\alpha,\gamma}(p',q;E)G_\gamma(q,E)\Gamma_{\gamma,j}(q) 
\label{eq:dressf} \\
\Sigma(E)_{i,j} & = & \int_C dq q^2 \sum_\gamma
\Gamma_{\gamma,i}(q)G_\gamma(q,E)\bar{\Gamma}_{\gamma,j}(q).
\label{eq:selfe}
\end{eqnarray}
where $\Gamma_{\alpha,i}(p)$ defines the coupling of the
$i$-th bare $N^*$ state to channel $\alpha$.

To search for nucleon resonances, we have developed\cite{ssl09}
 an analytic continuation method to find poles of 
 the scattering amplitude
$T_{\alpha,\beta}(p',p;E)$ on
 the unphysical sheets of the complex energy $E$-plane.
For multi-channel reactions considered here,  
there can be many poles associated
with a single resonance on  different unphysical sheets.
The pole  nearest to the physical sheet
is supposed to play a dominant effect on physical observables, and is called
the resonance pole traditionally. The other poles are called shadow poles.

\begin{figure}
\begin{center}
\includegraphics[width=12cm]{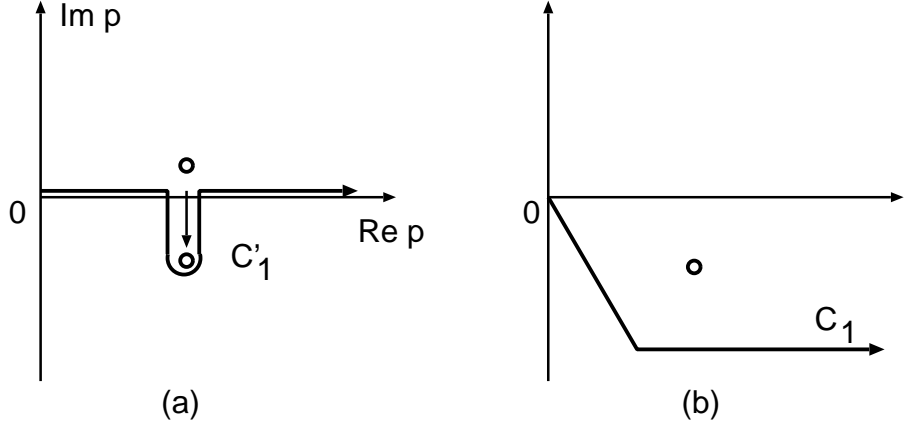}
\caption{The shift of the on-shell momentum (open circle) of the 
two-particle Green function Eq. (\ref{eq-g-st}) as energy E
moves from a real value above the threshold energy 
to a complex value with negative imaginary part.
 $C^\prime_1$ in (a) or $C_1$ in (b) is  the integration path 
for calculating Eqs.(\ref{eq:mext})-(\ref{eq:selfe})
amplitude for E on the unphysical plane.}
\label{fig:path}
\end{center}
\end{figure}

Since $v_{\alpha,\beta}$ and the bare vertex $\Gamma_{\alpha,i}$
are energy independent within the MSL formulation,
the analytic structure of the scattering
amplitude defined above as a function of $E$ is mainly determined by the
the Green functions  $G_\gamma(q,E)$.
Thus the key for selecting the amplitude on physical sheet or
unphysical sheet
is to take an appropriate path of momentum integration $C$ in
Eqs.(\ref{eq:fullt})-(\ref{eq:selfe}) according to
the locations of the singularities of the meson-baryon
Green functions $G_{\alpha}(p,E)$ as $E$ move to complex plane.
This can be done independently for each meson-baryon channel.
For  channel with stable particles such as $\pi N$ and $\eta N$,
the meson-baryon Green function is 
\begin{eqnarray}
G_{MB}(E,p) = \frac{1}{E - E_M(p) - E_B(p)} \,, \label{eq-g-st}
\end{eqnarray}
which has a pole at the on-shell momentum $p_0$ defined by
\begin{eqnarray}
E=\sqrt{m_M^2+p^2_{0}} + \sqrt{m_B^2+p^2_{0}}.
\label{eq-on-sh}
\end{eqnarray}
As an example, let us consider the analytic continuation
of the amplitude to the unphysical sheet of the $MB$ channel
when the energy $E$ is above the threshold  $Re(E)> m_B + m_M$
and $Im(E)<0$.
The on-shell momentum $p_0$ for such a $E$
is on the second and the fourth quadrant
of the complex momentum plane.
As $Im(E)$ becomes more negative 
as illustrated in Fig. \ref{fig:path}, the on-shell momentum (open
circle) moves into the fourth quadrant.
The amplitude on the unphysical sheet can be obtained by deforming
the path $C$ into  $C_1'$ or equivalently $C_1$
so that the on-shell momentum does not cross the integration contour.
We note here that for the energy below threshold ($Re(E)< m_B+m_M$)
the path $C_1$ will give amplitudes on the physical sheet.

\begin{figure}
\begin{center}
\includegraphics[width=6cm]{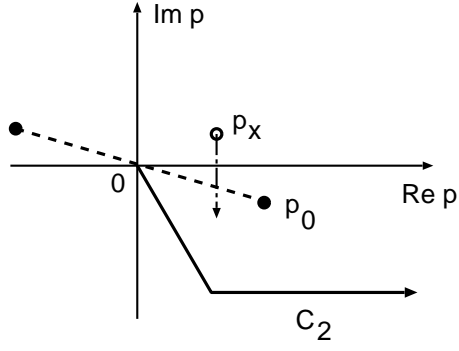}
\caption{ Contour $C_2$ for 
for calculating Eqs.(\ref{eq:mext})-(\ref{eq:selfe})
for E on the unphysical plane with the unstable particle
propagators, such as Eq.(\ref{eq-g-unst}) for $\pi \Delta$ channel.
See the text for the explanations of the dashed line and
the singularity $p_x$.}
\label{fig:c2}
\end{center}
\end{figure}

For the channels with unstable particle
 such as the $\pi \Delta$ as an example,
the Green function is of the following form
\begin{eqnarray}
G_{\pi\Delta}(E,p)  =  
\frac{1}{E- E_\pi(p) - E_\Delta(p)- \Sigma_{\Delta}(E,p)},
\label{eq-g-unst}
\end{eqnarray}
where
\begin{eqnarray}
\Sigma_\Delta(p,E)= \int_{C_3}  \frac{\{\Gamma_{\Delta,\pi N}(q)\}^2 q^2 dq}
{E-E_\pi(p)- [(E_\pi(q)+E_N(q))^2+p^2]^{1/2}}. \nonumber \\
\label{eq-sigma-pid}
\end{eqnarray}
The  $\pi\Delta$ Green function Eq. (\ref{eq-g-unst})
has a singularity at momentum $p=p_x$, which satisfies
\begin{eqnarray}
E-E_\pi(p_x)-E_\Delta(p_x)-\Sigma_\Delta(p_x,E)=0.
\label{eq-on-sh-pid}
\end{eqnarray}
Physically, this singularity corresponds to the $\pi\Delta$ two-body
'scattering state'.
There is also discontinuity of the
$\pi\Delta$ Green function  associated with the $\pi\pi N$ cut
in $\Sigma_\Delta$, as shown in the dashed line in Fig. \ref{fig:c2},
where $p_0$ is defined by  
\begin{eqnarray}
E=E_\pi(p_0)+[(m_\pi+m_N)^2+p^2_0]^{1/2}.
\end{eqnarray}
Therefore, for $Re(E)> m_B + m_M, 2m_\pi + m_N$,
the integration contour $C$ 
must be chosen to be below the $\pi\pi N$ cut (dashed line)
 and the singularity $p_x$, such
as the contour $C_2$ shown in  Fig.\ref{fig:c2},
for calculating amplitudes on the unphysical sheet.

\begin{figure}
\begin{center}
\includegraphics[width=6cm]{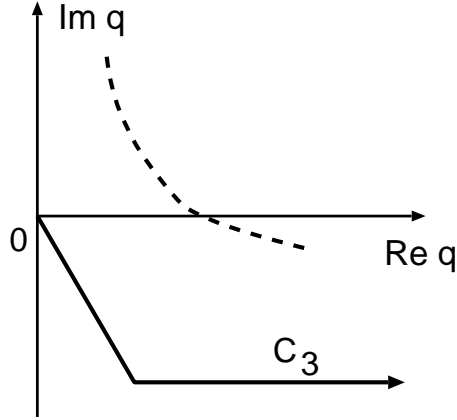}
\caption{ Contour $C_3$ for calculating the $\Delta$ self energy 
Eq.(\ref{eq-sigma-pid}) 
on the unphysical sheet. Dashed curve is the singularity $q_0$ of
the propagator in Eq. (\ref{eq:sing-q0}), which depends on
the spectator momentum $p$ on the contour $C_2$ of Fig.\ref{fig:c2}. }
\label{fig:c3}
\end{center}
\end{figure}

The singularity $q_0$ of the integrand of
Eq. (\ref{eq-sigma-pid})  depends on the spectator momentum $p$
\begin{eqnarray}
E-E_\pi(p) = [(E_\pi(q_0)+E_N(q_0))^2+p^2]^{1/2} \,.
\label{eq:sing-q0}
\end{eqnarray}
Thus  $q_0$ moves along the dashed curve, illustrated in Fig.\ref{fig:c3},
when the momentum $p$ varies along the path $C_2$ of Fig.\ref{fig:c2}.
To analytically continue $\Sigma_\Delta(p,E)$ to the unphysical sheet,
the contour $C_3$ of Eq. (\ref{eq-sigma-pid}) must be
 below $q_0$. A possible contour $C_3$ is the solid curve in Fig.\ref{fig:c3}.

\begin{figure}
\begin{center}
\includegraphics[width=10cm]{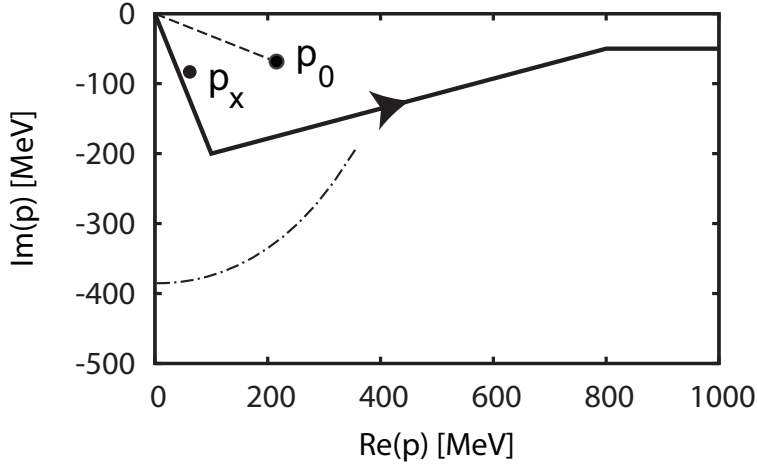}
\caption{The contour (solid curve) for calculating electromagnetic
matrix element. $p_0$ and $p_x$ are the singularities shown in 
Fig.\ref{fig:c2}. The dashed-dot curve is the singularity of the pion-exchange
$\gamma N \rightarrow \pi N$ matrix element at $E = 1357 - 76 i $ MeV.
\label{fig:actual-path} }
\end{center}
\end{figure}

We emphasize here that we can deform the contour $C$
only in the region where the potential $v_{\alpha,\beta}(p',p)$
and the bare $N^*$ vertex $\Gamma_{MB,N^*}(p)$ 
are analytic. The contours described
above only from considering the singularities of $MB$ and $\pi\pi N$
Green functions. Thus they must be modified according to the analytic structure
of the considered $v_{\alpha,\beta}(p',p)$ and $\Gamma_{MB,N^*}(p)$.
Within the MSL formulation, 
the t-channel meson exchange  potential
$v^t_{M'B',MB}(\vec{p}^{\,\,'},\vec{p})$
has  singularities at
\begin{eqnarray}
\Delta^2 - (\vec{p} - \vec{p}')^2 = 0
\end{eqnarray}
with $\Delta=E_{M'}(p')- E_M(p)$ or $E_{B'}(p')-E_B(p)$.
The form of $\Gamma_{MB,N^*}(p)$ is chosen such that its singularity is
at the pure imaginary momentum.
Thus the contours have to be chosen to also avoid these singularities.
As an example we show in Fig.\ref{fig:actual-path} 
the singularities associated with the $\pi\Delta$ channel at $E=1357 -76i$MeV.
The dotted line for $\pi\pi N$ cut
and the circle shows $p_X$ are the singularities from the Green's function,
as discussed above.
The most relevant singularity of the meson-exchange
potential in our investigation of electromagnetic
pion production amplitude is due to
the t-channel pion exchange of $\gamma N \rightarrow
\pi\Delta$,
which is shown as the dashed-dot curve. Thus the
integration contour has to be modified to the solid curve in  
Fig.\ref{fig:actual-path}.

\section{Extraction of resonance parameters}

The resonance energy ($M_R=M - i \Gamma/2$) is the position of
the pole of the scattering amplitude which is on the  sheet nearest
to the physical sheet.
In principle the resonance pole can be
found in the meson-exchange amplitude $t$ and/or
resonance amplitude $t^R$ of Eq. (\ref{eq:fullt}). 
Within the $\pi N$ model developed in Ref.\cite{jlms07} (JLMS),
we find\cite{sjklms09} that resonance poles are only from
 $t^R$. We therefore will only explain how
the residues of resonance poles are extracted from this term.

The poles  of $t^R$ are found from the zeros of the
determinant of $N^*$ propagator defined by Eq.(\ref{eq:tr-g})
\begin{eqnarray}
\Delta(E)= det[G_{N^*}^{-1}(E)]&=& 0 \,.
\label{eq:dete}
\end{eqnarray}
Near the resonance energy $M_R$, $N^*$ Green function
can be expressed as
\begin{eqnarray}
(G_{N^*}(E))_{ij} & = & \frac{\chi_i \chi_j}{E - M_R} \,,
\label{eq:pole-res}
\end{eqnarray}
where $i,j$ denote the bare $N^*$ state in the Hamiltonian and
$\chi_i$ represents $i$-th component of the dressed $N^*$
and satisfies
\begin{eqnarray}
\sum_j (G_{N^*}(M_R)^{-1})_{ij}\chi_j & = &
\sum_j [(M_R - m_{N^*_i})\delta_{ij} - \Sigma(M_R)_{ij}]\chi_j = 0.
\end{eqnarray}
If there is only one bare $N^*$ state, 
it is easy to see that
\begin{eqnarray}
\chi & = & \frac{1}{\sqrt{1 - \Sigma'(M_R)}} \,,
\end{eqnarray}
where $\Sigma'(M_R) = [d\Sigma/dE]_{E=M_R}$.
If we have two bare $N^*$ states, we find that
\begin{eqnarray}
\chi_1 & = & \sqrt{\frac{M_R - m_{N^*_2} - \Sigma_{22}(M_R)}{\Delta'(M_R)}}
\,, \\
\chi_2 & = & \frac{\Sigma_{12}(M_R)}{M_R - m_{N^*_2} \,, 
- \Sigma_{22}(M_R)}\chi_1
\end{eqnarray}
where $\Delta'(M_R)=[d\Delta/dE]_{E=M_R}$  can be evaluated using
Eq.(\ref{eq:dete}). 

We now examine how the residues $\chi_i$ can be used to see the effects
of resonance poles on the full  amplitude 
$T_{\alpha,\beta}$ defined by
Eq.(\ref{eq:fullt}).  
Near the resonance pole we can perform Laurent expansion 
of the on-shell amplitude
\begin{eqnarray}
T_{\alpha,\beta}(p^0_\alpha,p^0_\beta,M_R)
= \frac{\bar{\Gamma}^R_\alpha \bar{\Gamma}^R_\beta}{E - M_R}
 + C^0_{\alpha,\beta} + C^1_{\alpha,\beta}(E - M_R) + ... \,.
\label{eq:fullt-exp}
\end{eqnarray}
For the cases that the resonance poles are from $t^R$ term of 
Eq.(\ref{eq:fullt}), one can see from
the definitions Eqs.(\ref{eq:tr})-(\ref{eq:tr-g}) and 
Eq.(\ref{eq:pole-res}) that
 the vertex functions in the residue of
Eq.(\ref{eq:fullt-exp})
is determined by the dressed vertex $\bar{\Gamma}_{\alpha,j}$ defined by
Eq.(\ref{eq:dressf})
\begin{eqnarray}
\bar{\Gamma}^R_\alpha & = & \sum_j \chi_j 
\bar{\Gamma}_{\alpha,j}(p^0_\alpha,M_R).
\label{eq:gamma-r}
\end{eqnarray}
The terms $C^0_{\alpha,\beta}$ and $C^1_{\alpha,\beta}$ in 
Eq.(\ref{eq:fullt-exp}) also depend on the matrix elements of
meson-exchange amplitude $t$ of Eq.(\ref{eq:fullt}), 
but will not be discussed here.

The residue of the $\pi N$ elastic scattering amplitude $F_{\pi N,\pi N}$  
characterizes
the strength of the coupling of the resonance with $\pi N$
channel. Using the standard notation, we have at near resonance position
$M_R$
\begin{eqnarray}
F_{\pi N,\pi N} (E) & = & \frac{S_{\pi N, \pi N} (E) - 1}{2i}  
 = [\frac{R e^{i \phi}}{M_R - E}]_{E\rightarrow M_R} \,,
\end{eqnarray}
where $S_{\pi N,\pi N}$ is the partial-wave S-matrix. In terms of the 
normalization of JLMS model, we have
\begin{eqnarray}
F_{\pi N,\pi N} (M_R) = -
\pi \frac{p_0 E_N(p_0)E_\pi(p_0)}{M_R}T_{\pi N,\pi N}(p_0,p_0,M_R)
\end{eqnarray}
We thus have
\begin{eqnarray}
R e^{i \phi} & = & 
\pi \frac{p_0 E_N(p_0)E_\pi(p_0)}{M_R} \bar{\Gamma}^R_{\pi N}
\bar{\Gamma}^R_{\pi N} \,.
\label{eq:pin-resid}
\end{eqnarray}
The elasticity  of a resonance is then defined as
\begin{eqnarray}
\eta_e= \frac{R}{-Im (M_R)}
\label{eq:elasticity}
\end{eqnarray} 

The electromagnetic $N$-$N^*$ transition form factor 
is defined by matrix element of the electromagnetic
currents between nucleon and $N^*$
\begin{eqnarray}
A_{3/2}(Q^2) & = & X <N^*,s_z=3/2|-\vec{J}(Q^2)\cdot\vec{\epsilon}_{+1}|N,s_N=1/2>  \label{a3} \,, \\
A_{1/2}(Q^2) & = & X <N^*,s_z=1/2|-\vec{J}(Q^2)\cdot\vec{\epsilon}_{+1}|N,s_N=-1/2> \label{a1}\ ,,\\
S_{1/2}(Q^2) & = & X <N^*,s_z=1/2| J^0(Q^2)|N,s_N = 1/2> \label{s1} \,,
\end{eqnarray}
where $\vec{\epsilon}_{+1}  =  - (\hat{x} + i \hat{y})/\sqrt{2}$, 
\begin{eqnarray}
X       & = & \sqrt{\frac{E_N(\vec{q})}{m_N }} \frac{1}{\sqrt{2K}}\,, \nonumber
\end{eqnarray}
with $K  =  (M_R^2 - m_N^2)/(2M_R)$. The above definition was originally
introduced for the constituent quark model\cite{copley}.
If $<N^*|$ is a resonance state, then the above expression at resonance
position $M_R$ must be evaluated by using Eq.(\ref{eq:gamma-r}).
We thus have
\begin{eqnarray}
A_{3/2}(Q^2) & = & X X' \sum_j \chi_j \bar{\Gamma}^R_{\gamma^*
 N,j}(Q^2,M_R,\lambda_\gamma=1,\lambda_N=-1/2) \, \label{eq:emff},\\
\end{eqnarray}
where
\begin{eqnarray}
X' & = & \sqrt{\frac{(2j+1)(2\pi)^3(2q_0)}{4\pi}} \,. \nonumber
\end{eqnarray}
Here the additional 
factor $X'$ is due to our normalization of the vertex function
$\bar{\Gamma}$.
The above helicity amplitudes are in general complex number.
$A_{1/2}$ and $S_{1/2}$ have  similar expressions.

\section{Illustrative Results and Discussions}

In this section, we illustrate our procedures by presenting
the results for the pronounced  resonances
in $P_{33}$, $D_{13}$ and in the most complex $P_{11}$ partial waves. 
Their pole positions were
determined in Ref.\cite{sjklms09} and are listed in Table \ref{tab:poles}.
Our results for $P_{33}(1232)$
and $D_{13}(1520)$ agree well with the
values listed by Particle Data Group\cite{pdg} (PDG). 
For $P_{11}$ channel we found three poles below
2 GeV.  Two of them near 1360 MeV are close to the $\pi\Delta$ threshold.
This finding is consistent with the  earlier analysis of VPI\cite{vpi84} and
Cutkosky and Wang\cite{cut}, and the recent analysis by the
GWU/VPI\cite{gwu-vpi} 
and Juelich\cite{juelich} groups.
We have shown in Ref. \cite{sjklms09} that all of the  three $P_{11}$
resonances listed in Table \ref{tab:poles} correspond to
a single bare $N^*$ state at 1736 MeV. This is a dynamical verification of the
resonance pole-shadow pole relation in coupled-channels reactions, 
as discussed by Eden and
Taylor\cite{et63}, Kato\cite{kato65}, and Morgan and Pennington\cite{mp87}. 

The extracted residues $Re^{i\phi}$, defined in Eq.(\ref{eq:pin-resid}),
 for $\pi N$ amplitude 
 are compared with some of the previous works in Table 
\ref{tab:residues-pin}. We see that the agreement in $P_{33}$ and $D_{13}$
are excellent. For $P_{11}$, there are significant differences between
four analysis. As discussed in Ref.\cite{cut}, it could be mainly
due to the differences in the employed reaction models. On the other hand,
the difference between the predicted $P_{11}$ amplitudes at 
$W > $ about 1.6 GeV could be the reason why the third
$P_{11}$ pole is not found in Juelich analysis.

 From the values of $M_R$ of Table \ref{tab:poles} and  $R$ of Table 
\ref{tab:residues-pin}, we can evaluate 
the elasticities $\eta_e$ using Eq.(\ref{eq:elasticity}).
The results 
are also listed in Table \ref{tab:poles}. We see that our results
agree well with PDG values, while some investigations are needed to
 understand better the comparisons for the
two $P_{11}$ poles near 1360 MeV which are close to $\pi\Delta$ threshold.

To extract helicity amplitudes using Eq.(\ref{eq:emff}), we use the
multipole amplitudes calculated from using the parameters determined in
Ref.\cite{jklmss09}. 
Our results
at photon point are listed in Table \ref{tab:residues-gn}.
We observe that the real parts
of our results for $P_{33}$ and $D_{13}$ are in good agreement with
several previous results\cite{arndt04,ahrens04,dugger07,blanpied01}.
The large differences in $P_{11}$ indicate that more investigations are needed
to understand the differences between our resonance extraction method  within
a coupled-channel model and other methods which are mainly based on
the Briet-Wigner parametrization of single channel
K-matrix amplitudes.

\begin{figure}[th]
\centering
\includegraphics[width=8cm]{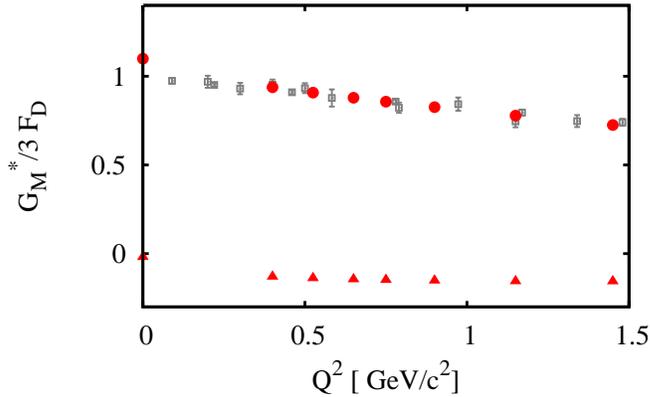}
\caption{The magnetic $N$-$\Delta$ (1232) transition form factor
$G^*_M(Q^2)$ defined in Ref.\cite{sl96}. $G_D=1./(1+Q^2/b^2)^2$ with
$b^2 = 0.71$ (GeV/c)$^2$. The solid circles (solid triangles) are 
the real (imaginary) parts of our results. The other data points are
 from previous analysis\cite{ndff-data}. }
\label{fig:p33-ff}
\end{figure}

\begin{figure}[th]
\centering
\includegraphics[width=5cm]{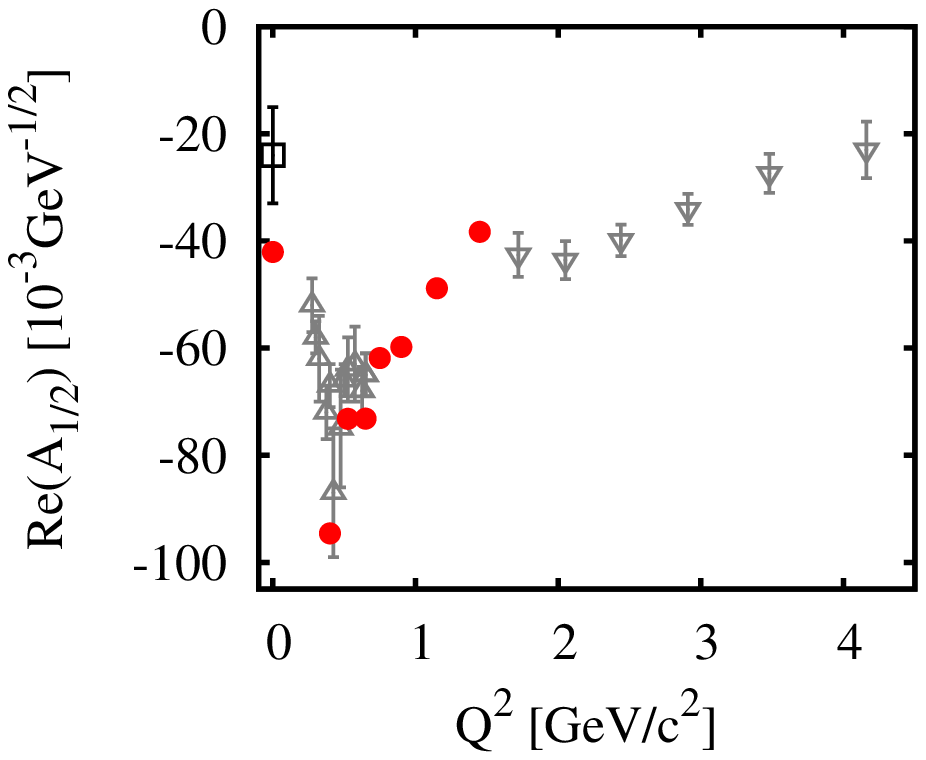}
\includegraphics[width=5cm]{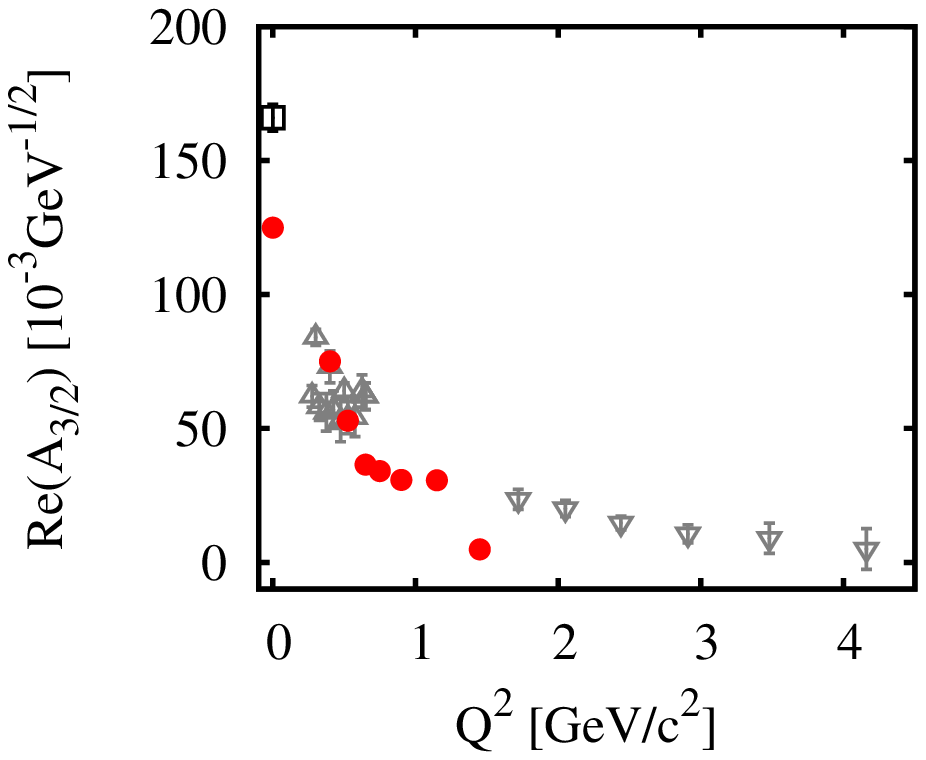}
\includegraphics[width=5cm]{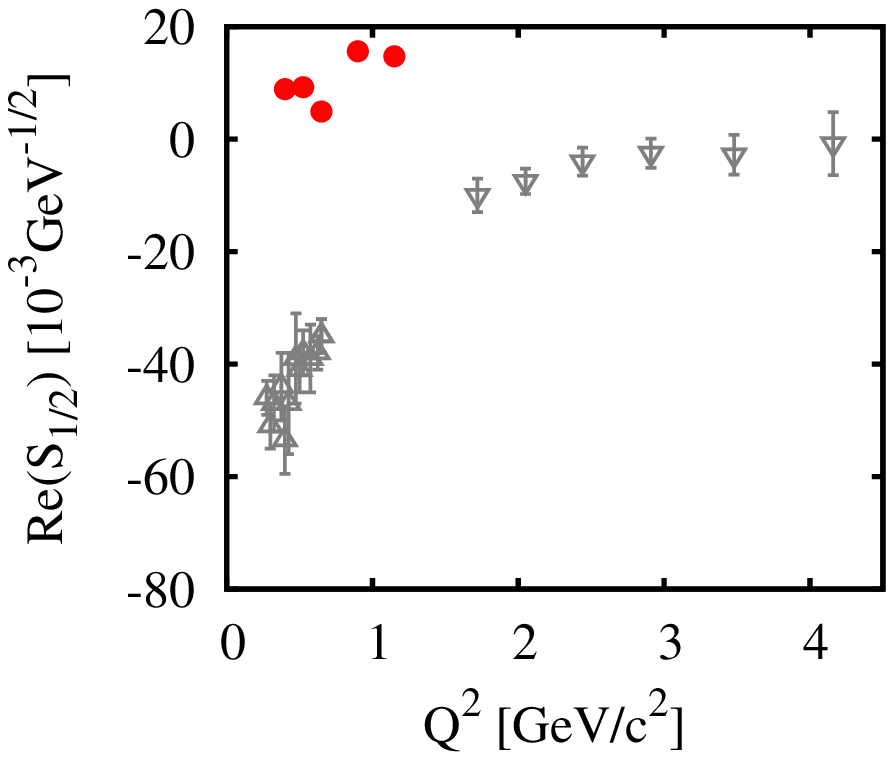}
\caption{The solid circles are the real parts of
the extracted $\gamma N \rightarrow N^*(D_{13}(1520))$ form factors.
The data are from CLAS collaboration\cite{inna}.
transition form factors}
\label{fig:d13-ff}
\end{figure}

For $P_{33}$ we can use the standard relation\cite{sl96} to
evaluate the $N$-$\Delta$ magnetic transition form factor $G_M^*$
in terms of helicity amplitudes. The real parts of
our results are the solid circles
in Fig.\ref{fig:p33-ff}, which are in good agreement with the
previous analysis. In the same figure, we also show that the imaginary parts
of our results are much weaker. This result and the results of 
Table \ref{tab:residues-gn} suggest that we can only make meaningful 
comparisons with the results from analysis based on the
Briet-Wigner 
parametrization of single channel K-matrix amplitudes only for
the cases that the imaginary parts are small.
This turns out to be also the case of the $D_{13}(1521)$ resonance.
In Fig.\ref{fig:d13-ff}, we see that the real parts of our $A_{3/2}$
and $A_{1/2}$ are
in good agreement with the results from CLAS collaboration\cite{inna}. 
The large 
differences in $S_{1/2}$ perhaps are mainly from the fact that
the longitudinal parts of the amplitudes can not be well
determined with the available data.

\begin{figure}[th]
\centering
\includegraphics[width=5cm]{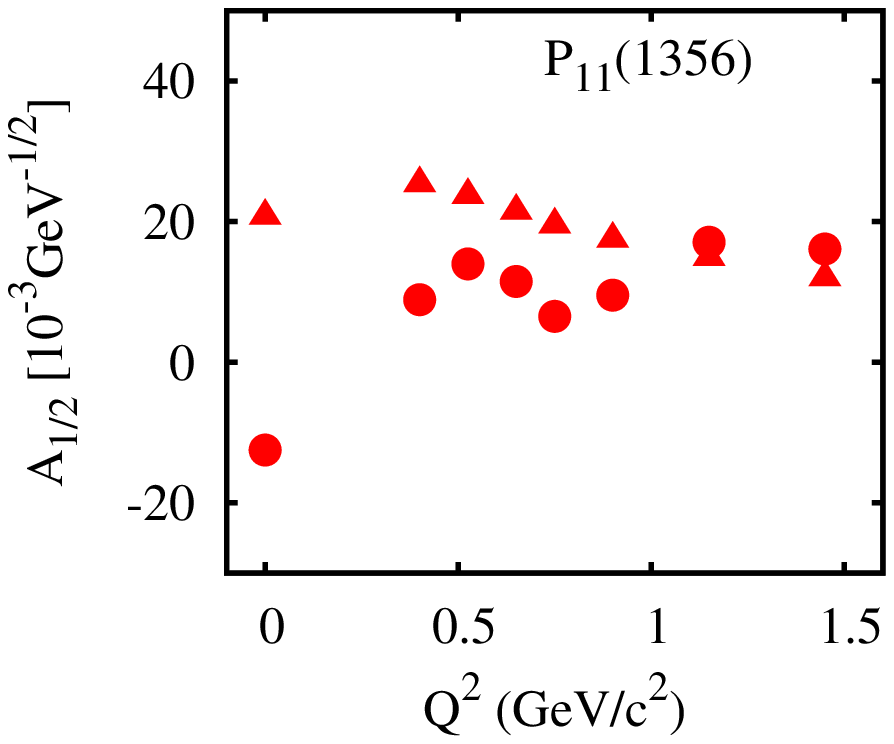}
\includegraphics[width=5cm]{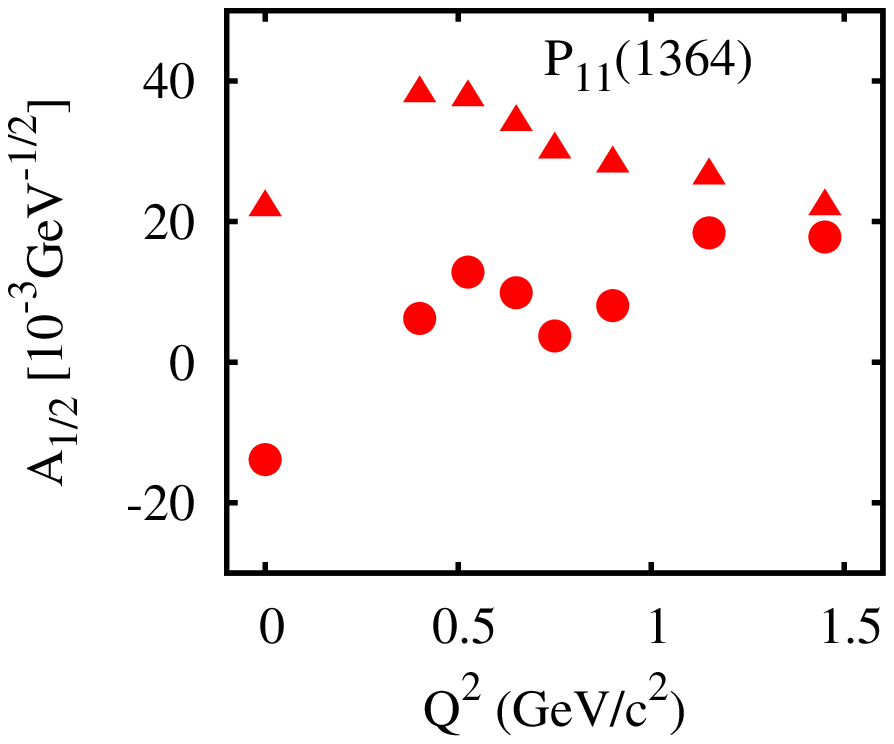}
\includegraphics[width=5cm]{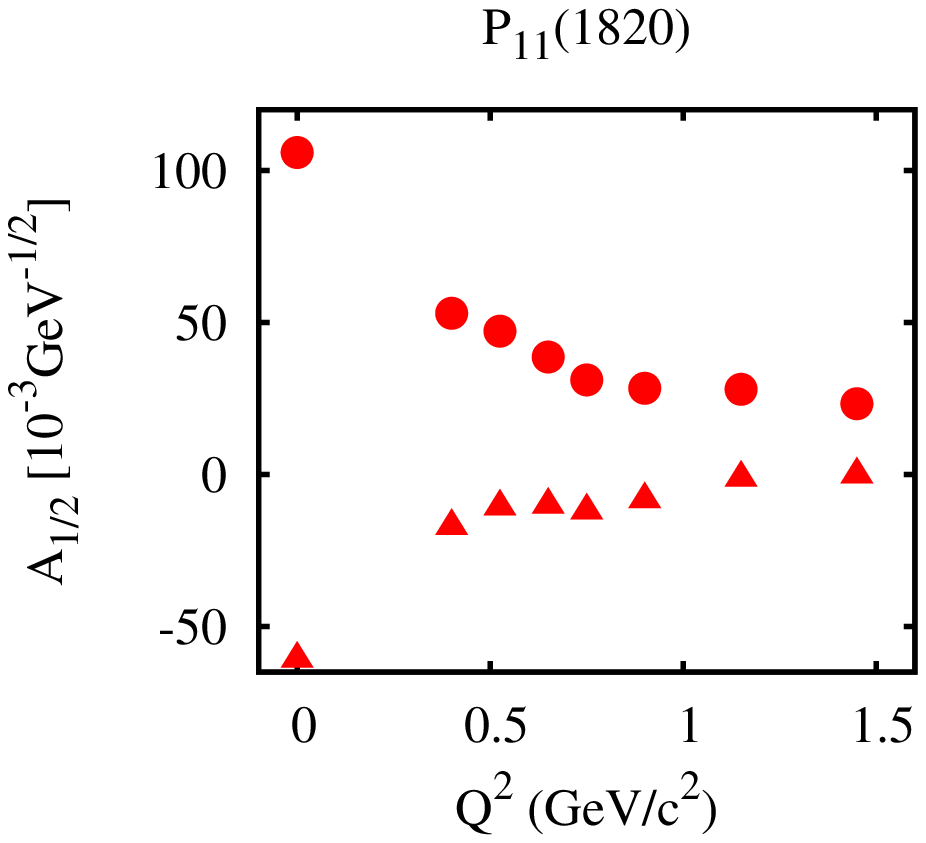}
\caption{The extracted 
$\gamma N \rightarrow N* (1356),  N* (1364), N* (1820))$
transition form factors of $P_{11}$. The solid circles (solid triangles) are
their real (imaginary) parts.  }
\label{fig:p11-ff}
\end{figure}

For $P_{11}$, the imaginary parts of the calculated helicity amplitudes
for the three poles listed in Table \ref{tab:poles}
are very large. Thus it is not clear how to compare our results
with previous results. We thus
show both the real parts (solid circles) and imaginary
parts (solid triangles)
in Fig. \ref{fig:p11-ff}.
It seems that the structure of $N^*(1356)$ and $N^*(1364)$ are similar.
In particular their real parts of  $A_{1/2}$ change sign at low $Q^2$, similar
to what haven been seen in the results from CLAS collaboration\cite{inna}.
However, because of the double pole structure and the
large imaginary parts, more detailed
investigations are need to make meaningful comparison with previous results.

\section{summary}
 
In this paper, we have briefly reviewed
 the analytic continuation method developed
in Ref.\cite{ssl09} and explained how it is used to determine the residues of
nucleon resonance poles. To illustrate our method,
we have  presented the results for
resonances in $P_{33}$, $D_{13}$, and $P_{11}$ partial waves.

For residues associated with $\pi N$ channel, we agree with most
of the previous results\cite{vpi84,cut,gwu-vpi,juelich}
 for $P_{33}(1232)$, $D_{13}(1521)$ and 
two $P_{11}$ poles near 1360 MeV. For $P_{11}(1820)$, the calculated
elasticity $\sim 8 \%$  agree well with the value $ \sim 10-20 \%$
of PDG and Ref.\cite{cut},  
while this resonance is not reported in
the analysis of Refs.\cite{vpi84,gwu-vpi,juelich}.

For residues associated with $\gamma N$ channel, the corresponding helicity 
amplitudes for $P_{33}$ (1232) and $D_{13}$ (1521)
 are dominated by their real parts which are in good agreement with other
analysis based on the Briet-Wigner parametrization of K-matrix amplitudes.
For $P_{11}$ resonances, the extracted helicities amplitudes have large
imaginary parts and more investigations are needed to compare our results
with previous analysis. 

Our next necessary task is to examine how to define the residues associated
with unstable $\pi\Delta$, $\rho N$, and $\sigma N$ channels.
Our effort in this direction along with our complete results for
the 14 nucleon
resonances extracted in Ref.\cite{sjklms09} will be reported elsewhere.

\begin{table}[t]
\caption{Resonance poles ($Re M_R, -Im M_R)$ MeV and
elasticity $\eta_e$ (Eq.(\ref{eq:elasticity})) extracted in Ref.\cite{kjlms09}
 }
\begin{tabular}{|c|c|c|c|c|c|c|c|c|c|}
\hline
  &          & $M_R$ (EBAC-DCC)  &     $M_R$ (PDG) & & $\eta_e$ (EBAC-DCC)&
$\eta_e$ (PDG) \\
\hline
$P_{33}$  & & (1211, 50) &   (1209 - 1211,    49 - 51)& & 100$\%$  &  100 $\%$ \\
\hline
\hline
$D_{13}$  &  & (1521, 58)     & (1505 - 1515   ,  52   - 60)& & 65 $\%$ 
& 55 - 65 $\%$\\
\hline
\hline
$P_{11}$ &   & (1357, 76)    & (1350 - 1380, 80 - 110) & & 49 $\%$ & 60 - 70  $\%$\\

         &   & (1364,105)   &                        &  &  61 $\%$ & \\

          &  & (1820, 248),   & (1670 - 1770, 40 - 190) & & 8 $\%$ & 10 - 20 $\%$ \\
\hline
\end{tabular}
\label{tab:poles}
\end{table}

 \begin{table}[t]
\caption{The extracted $\pi N$ residues $R e^{i\phi}$ defined by 
Eq.(\ref{eq:pin-resid}.}
 \begin{tabular}{|c|cc|cc|cc|cc|}
\hline
 & EBAC-DCC &        & GWU-VPI\cite{gwu-vpi}  &    & Cutkosky\cite{cut}& 
&Juelich\cite{juelich} & \\
 & R & $\phi$ & R & $\phi$ & R & $\phi$ &R & $\phi$  \\
 \hline
 $P_{33}(1210)$ & 52& -46  &  52&  -47    &  $53$ & $-47$& 47 & -37
  \\
 \hline
 $D_{13}(1521)$ &  38&  7  & 38 & -6    &  $35$  &$-12$ & 32 & -18 \\
 \hline
 $P_{11}(1356)$ & 37  & -111  &  38  & -98    &  $52$ & $-100$& 48 & -64\\
 $\,\,\,\,\,\,\,\,(1364)$ &   64  & -99  & 86 & -46    &  - & -& & \\
 $\,\,\,\,\,\,\,\,(1820)$ &   20 & -168  & - & -    & 9   & -167 &-&- \\
 \hline
 \end{tabular}
\label{tab:residues-pin}
 \end{table}

 \begin{table}[t]
\caption{The extracted $\gamma N \rightarrow N^*$ helicity amplitudes 
are compared with previous results.}
 \begin{tabular}{|c|c|c|c|c|c|c|}
\hline
              &         & EBAC     & Arndt04/96 & Ahrens04/02  &  Dugger07 & Blanpied01 \\
\hline
 $P_{33}(1210)$  & $A_{3/2}$ & -269+12i & -258      &   -243  &     & -267 \\
                 & $A_{1/2}$ & -132+38i & -137      &   -129  &     &  -136\\
\hline
 $D_{13}(1521)$  & $A_{3/2}$ & 125+22i & $165\pm 5$     &   $147\pm 10$ & 
$142\pm 2$ &\\
                 & $A_{1/2}$ & -42+8i  & $-20\pm 73$   &   $-28\pm 3 $ & 
$ -28\pm 2$ &\\
\hline
$P_{11}(1356)$      & $A_{1/2}$ & -12+2i & $-63\pm 5$     &    & $-51\pm 2$ &\\
$\,\,\,\,\,\,\,\,(1364)$      & $A_{1/2}$ & -14+22i &                &    & & \\
\hline
 \end{tabular}
\label{tab:residues-gn}
 \end{table}

\begin{thebibliography}{90}

\bibitem{ssl09}
N. Suzuki, T. Sato and T. -S. H, Lee,
Phys. Rev. C{\bf 79}, 025205 (2009).

\bibitem{msl07}
A. Matsuyama, T. Sato, and T. -S. H. Lee,  Phys. Rept. {\bf 439}, 193 (2007).

\bibitem{sjklms09}
N. Suzuki, B. Julia-Diaz, H. Kamano, T.-S. H. Lee, A. Matsuyama,
and T. Sato,
arXiv:0909.1356[ncl-th], submitted to Phys. Rev. Lett (2009).


\bibitem{jlms07}
B. Julia-Diaz, T. -S. H. Lee, A. Matsuyama, and T. Sato,
Phys. Rev. C{\bf 76},  065201 (2007).

\bibitem{kjlms09}
H. Kamano, B. Julia-Diaz, T. -S. H.  Lee, A. Matsuyama, and T. Sato,
Phys. Rev. C{\bf 79},  025206 (2009);
H. Kamano, B. Julia-Diaz, T. -S. H.  Lee, A. Matsuyama, and T. Sato,
arXiv:0909.1129 [nucl-th].


\bibitem{jlmss08}
B. Julia-Diaz, T. -S. H. Lee, A. Matsuyama, T. Sato and L. C. Smith,
Phys. Rev. C{\bf 77}, 045205 (2008).

\bibitem{jklmss09}
B. Julia-Diaz, H. Kamano, T. -S. H. Lee, A. Matsuyama, T. Sato
and N. Suzuki, Phys. Rev. C{\bf 80}, 025207 (2009).


\bibitem{copley}
L. A. Copley, G. Karl, and E. Obryk
Nucl. Phys. {\bf B13},  303 (1969).


\bibitem{pdg}
C. Amsler et al., Phys. Lett. {\bf B667}, 1 (2008).

\bibitem{vpi84}
R.A. Arndt, J. M. Ford, L. D. Roper, Phys. Rev. D{\bf 32}, 1085 (1985).

\bibitem{cut}
R.E. Cutkosky and S. Wang, Phys. Rev. D. {\bf 42}, 235 (1990);
R. E. Cutkosky, C. P. Forsyth, R. E. Hendrick and R. L. Kelly,
Phys. Rev. D{\bf 20}, 2839 (1979).

\bibitem{gwu-vpi}
R. A. Arndt, W. J. Briscoe, I. I. Strakovsky, and R. L. Workman,
Phys. Rev C{\bf 74}, 45205 (2006).

\bibitem{juelich}
D\"{o}ring M, Hanhardt C, Huang F, Krewald S and Mei\ss ner U -G,
arXiv:0903.1781 [nucl-th];
D\"{o}ring M, Hanhardt C, Huang F, Krewald S and Mei\ss ner U -G,
Nucl. Phys. {\bf A829}, 170 (2009).

\bibitem{et63}
R. J. Eden and J. R. Taylor, Phys. Rev. Lett. {\bf 11}, 516 (1963).

\bibitem{kato65}
M. Kato, Ann. Phys. (N.Y.) {\bf 31}, 130 (1965).

\bibitem{mp87}
D. Morgan and M.R. Pennington, Phys. Rev. Lett. {\bf 59}, 2818 (1987).


\bibitem{arndt04}
R. A. Arndt, W. J. Briscoe, I. I. Strakovsky, and
R. L. Workman, Phys. Rev. C{\bf 66}, 055213 (2002);
R. A. Arndt, I. I. Strakovsky, and R. L. Workman,
Phys. Rev. C{\bf 53}, 430 (1996).


\bibitem{ahrens04}
J. Ahrens et al., Eur. Phys. J. A{\bf 21}, 323 (2004);
J. Ahrens et al., Phys. Rev. Lett, {\bf 88}, 232002 (2002).

\bibitem{dugger07}
M. Dugger et al., Phys. Rev. C{\bf 76}, 025211 (2007).


\bibitem{blanpied01}
G. Blanpied et al., Phys. Rev. C{\bf 64}, 025203 (2001).


\bibitem{sl96}
T. Sato and T. -S. H. Lee, Phys. Rev C{\bf 54}, 2660  (1996).

\bibitem{ndff-data}
W. Bartel et al.,Phys. Lett {\bf 28B}, 148 (1968);
K. B\"{a}tzner et al., Phys. Lett. {\bf 39B}, 575 (1972);
J. C. Alder et al., Nucl. Phys. {\bf B46}, 573  (1972);
S. Sterin et al., Phys. Rev. D{\bf 12}, 1884 (1975).


\bibitem{inna}
G. Aznauryan,V,D. Burkert, et al. (CLAS Collaboration), arXiv:0909.2349v2;
V.I. Mokeev, V.D. Burkert et al. arXiv:0906.4081.
\end{thebibliography}
\end{document}